# On the Structural Origin of the Catalytic Properties of Inherently Strained Ultrasmall Decahedral Gold Nanoparticles

Michael J. Walsh,[†] Kenta Yoshida,[∥] Akihide Kuwabara,[∥] Mungo L. Pay, Pratibha L. Gai,[*,†,‡] and Edward D. Boyes[†,]

[†]Departments of Physics, [‡]Chemistry and Electronics, University of York, The York JEOL Nanocentre, York, YO10 5DD, United Kingdom

[∥]Nanostructures Research Laboratory, Japan Fine Ceramics Center, 2-4-1 Mutsuno, Atsuta-ku, Nagoya, 456-8587, Japan

*s Supporting Information

**ABSTRACT:** A new mechanism for reactivity of multiply twinned gold nanoparticles resulting from their inherently strained structure provides a further explanation of the surprising catalytic activity of small gold nanoparticles. Atomic defect structural studies of surface strains and quantitative analysis of atomic column displacements in the decahedral structure observed by aberration corrected transmission electron microscopy reveal an average 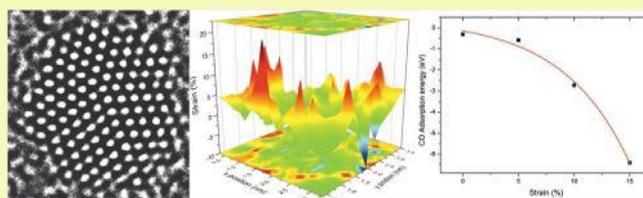 expansion of surface nearest neighbor distances of 5.6%, with many strained by more than 10%. Density functional theory calculations of the resulting modified gold d-band states predict significantly enhanced activity for carbon monoxide oxidation. The new insights have important implications for the applications of nanoparticles in chemical process technology, including for heterogeneous catalysis.

The application of catalytic nanoparticles promises to play a crucial role in the pursuit of environmentally sustainable and economically viable energy solutions. Gold nanoparticles less than 5 nm in size are known to be highly active catalysts, particularly for the low-temperature oxidation of carbon monoxide.[1,2] While the origin of gold's surprising activity has been described as being due to quantum size effects,[3,4] perimeter sites[5] and particle-support interfaces,[6] the role of low-coordinated surface atoms[7−11] has perhaps received the most support with theoretical and experimental evidence showing stronger binding on kink or stepped sites as opposed to planar surfaces. This suggests that particle morphology may be critical for catalysis.

The effect of reducing the coordination number of atoms modifies the band structure, shifting the d-states closer to the Fermi level. This effect is explained by the d-band model,[12] in which a reduced overlap of the wave functions leads to a narrowing of the d-band and a subsequent increase in the band's population. The resulting upshift of the d-band center increases the strength of the surface-molecule interaction, as fewer of the antibonding states are occupied.[13] In this model, a reduction in coordination number and an increase in tensile strain have the same effect on the d-band of transition metals.

DFT calculations by Mavrikakis et al.[14] predict considerable variation in adsorption energies with strain for CO and O on Ru and that as the lattice constant is increased chemisorption becomes stronger. Furthermore, Gsell et al. provided direct evidence of oxygen being preferentially adsorbed on areas of Ru that have been strained by subsurface Ar bubbles,[15] while later studies have shown stronger binding of adsorbates on layers of Pd, Pt, and Ni that are strained as a result of a lattice mismatch with the support.[16−18] For Au supported on $TiO_2$, strain induced by the support material has been observed[19] with further DFT calculations showing such support-induced strain would significantly enhance the adsorption of O and $O_2$ on gold.[20]

Despite the promising catalytic properties of surfaces with support-induced strain, their application on a large scale is unlikely, as the strain will be released by the formation of dislocations or reduced by annealing at elevated operating temperatures. Furthermore, the strain will be largest at the support interface and therefore not easily accessible to incoming gas molecules. Gold nanoparticles in the size range of 2−10 nm are typically multiply twinned with the most stable forms being the non-space filling icosahedral and decahedral structures.[21−23] In the decahedral case, the particle is made up of five tetrahedra bound by (111) twin planes. As the angle between each (111) plane should be 70.53°, there is a missing angle of 7.35°, meaning such structures are crystallographically forbidden. However, the associated reduction in surface energy allows for the necessary internal strain to be accommodated,[24] while re-entrant facets may further reduce the particles surface energy.[25] Further work on the inherent strain in decahedral gold nanoparticles by Johnson et al. using geometrical phase analysis (GPA) described the presence of both a disclination and shear gradients as well as the important effect of elastic anisotropy.[26]

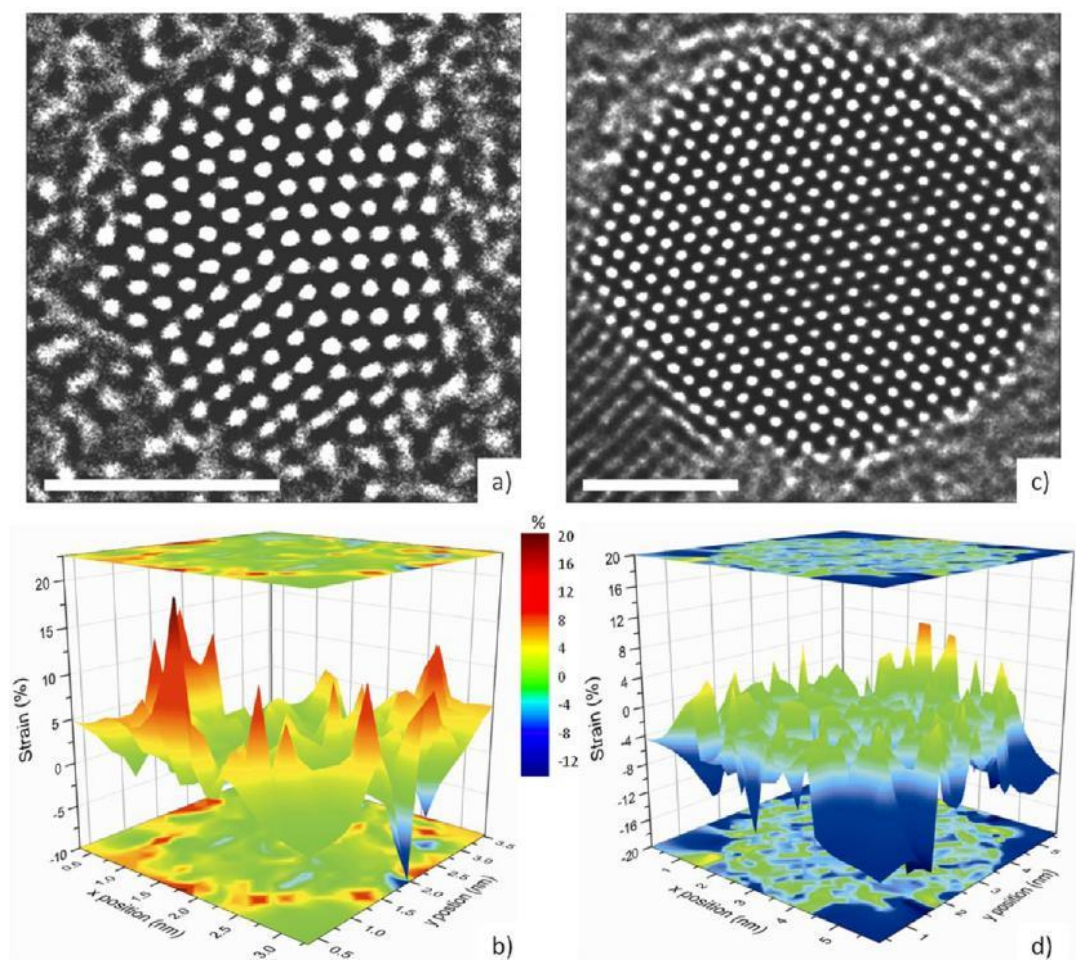

Figure 1. AC-TEM images of a decahedral particle (a) and single crystal particle (c) with corresponding strain distributions (b,d). Scale markers indicate 2 nm.

In the present work, we have used aberration-corrected transmission electron microscopy (AC-TEM) to experimentally observe and analyze the structure of gold nanoparticles in the size range for which gold is chemically active. The use of aberration correction provides a consistent phase contrast throughout the image, while operation at close to zero defocus minimizes image delocalization, allowing from a single image accurate observations of interatomic spacings, crystal structures, interfaces, and external surfaces. The use of a wider gap polepiece made possible with aberration correction provides space for a regular hot stage and an increased specimen tilt range.[27] Imaging was performed with a slightly negative spherical aberration value corrected to the third order and a slightly positive defocus value for optimum contrast, allowing for maximum precision in the image.[28]

Figure 1a shows an example of a gold decahedral particle in the [110] orientation supported on 3.5 nm thick amorphous carbon (preparation described in Supporting Information). The directly interpretable image provided by aberration correction allows for minimal exposure of the particle to the electron beam. The particle is approximately 3 nm in diameter and shows the classical Marks decahedral structure.[24] However, fully quantitative analysis of the local strain in multiply twinned structures by conventional approaches such as GPA is difficult, as the need to define a reference vector leads to discontinuities across the crystal boundaries, as shown in Supporting Information Figure S1.

Through an alternative approach working in real space, the intensity peak corresponding to each atomic column is identified, and by measuring each interatomic distance from one column to the next the need for a reference vector is removed, as illustrated in Figure 2. In order to accurately determine the position of each atomic column a high quality spherical aberration free image with consistent phase contrast and minimal delocalization is essential. The particle must also be orientated in a zone axis in order for clear resolution of the individual columns. From the initial aberration-corrected TEM image, each atomic column was first identified by putting a threshold on the image so that only the atomic columns were included. At this point, each atomic column highlighted was checked manually in order to ensure that all peaks had been identified; no random bright areas had been misidentified as atomic columns and there were no spurious peaks present as a result of intensity variations across the intensity profile of the peak. Plots of the peak intensity integrated across the atomic columns revealed the intensity distributions to be Gaussian and symmetrical about the center of the peak. Therefore the position of the atomic column was defined as the centroid of the selected intensity above the threshold value. The accuracy with which it is possible to define the atomic column coordinate is the largest source of error and determines the final uncertainty in the measured nearest neighbor distance. This is a random error due to the pixel size (0.01 nm) and leads to an uncertainty of ±3.5% in the measured nearest neighbor distance. Any strain resulting in a bending of the column is not directly observable and will lead to a slight averaging of the column position.

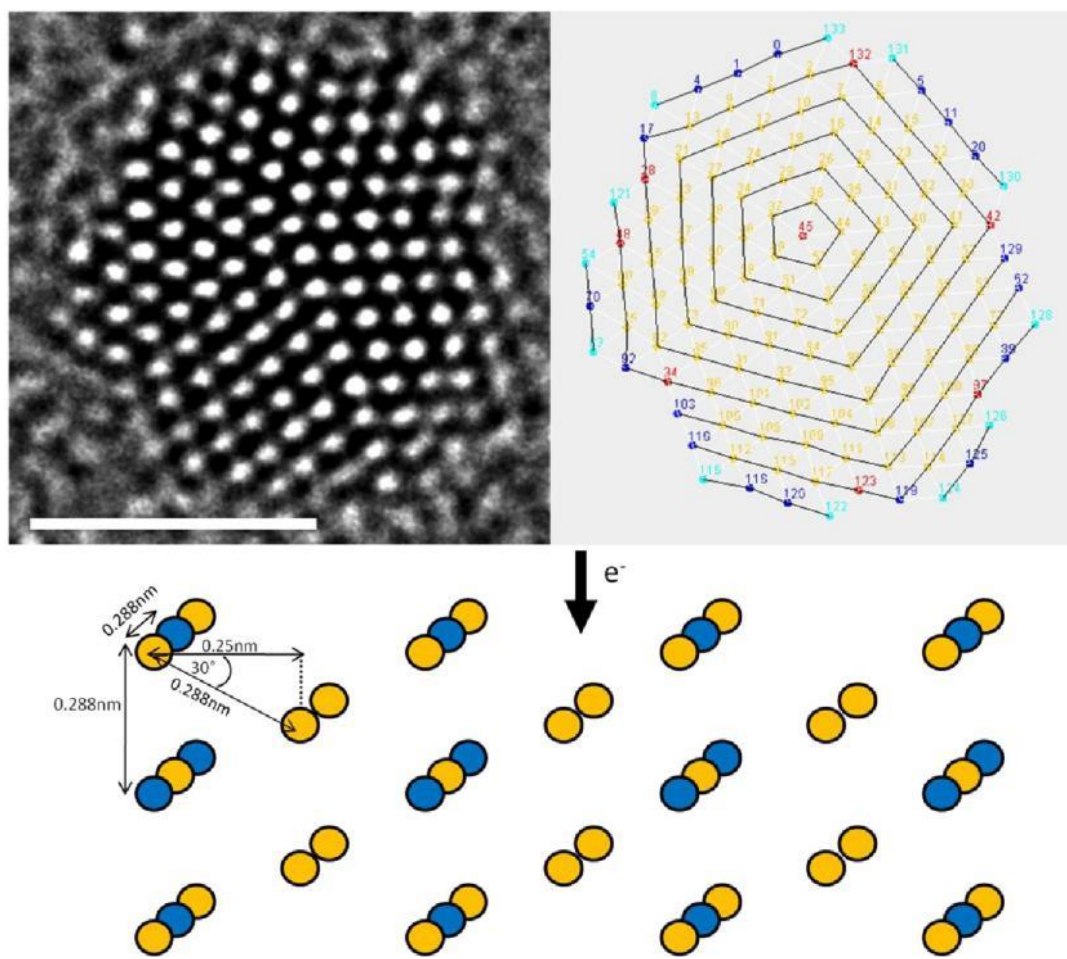

Figure 2. Atomic column displacements are measured by identifying the x−y coordinates of each column and measuring the distances from one coordinate to the next, correcting for crystallographic orientation.

From this point, the x−y coordinates of each column are fed into code developed in-house in order to provide the corresponding nearest neighbor distances. Understanding of the crystallography of the particle in question allows us to compare each distance measured to the known nearest neighbor distance of bulk gold (0.288 nm) and to analyze both the amount of strain with respect to the bulk value and its location. While the two-dimensional images only allow analysis of strain in the x−y plane, it can be assumed that there is minimal strain in the z-direction, as the disclination axis is aligned parallel to the direction of the electron beam. For the purpose of correcting measured distances in the x−y projection to real nearest neighbor distances it is assumed there is no strain in the axis parallel to the beam, while no such assumption is needed for interatomic distances measured between atoms in a plane normal to the beam, as depicted by the black lines in Figure 2.

From the analysis of Figure 1a, we can see that the measured atomic spacings follow a log-normal distribution (see Supporting Information Figure S2) with a mean nearest neighbor distance (with standard error) of 0.290 ± 0.002 nm. Figure 1b is a three-dimensional (3D) surface plot of the measured nearest neighbor distances and shows regular expansions of up to 15 ± 3.5%. These highly strained interatomic distances predominantly correspond to both circumferential and radial strain at the surface. Corresponding 2D strain maps are provided in Supporting Information Figure S3, and both these and Figure 1b show that when the need to define a reference vector is removed, the location of the twin boundaries is not observed, thus showing that the strain is coherent across the twin interface. The distribution of strain throughout the particle is described in Figure 3 by plotting the mean strain measured as a function of distance from the center of the particle. From this graph some compression near the crystallographic center is observed, while an average expansive strain of 5.6% is found at the surface. The large error bar corresponding to the average surface strain value indicates the larger spread of nearest neighbor distances measured at the surface. The d-band model suggests that changes in the local lattice parameter of this magnitude will have a profound influence on the band structure of the particle and thus on their physical and chemical properties. These effects are attributed to the inherent strain associated with the crystallographically forbidden nature of the structure and are larger than anything previously reported due to interface strains.[19] Accordingly they may be among the more important factors in practical catalytic applications since they are both the largest effects known to have been reported to date and they affect directly the free outer surfaces exposed for reactions with gas molecules.

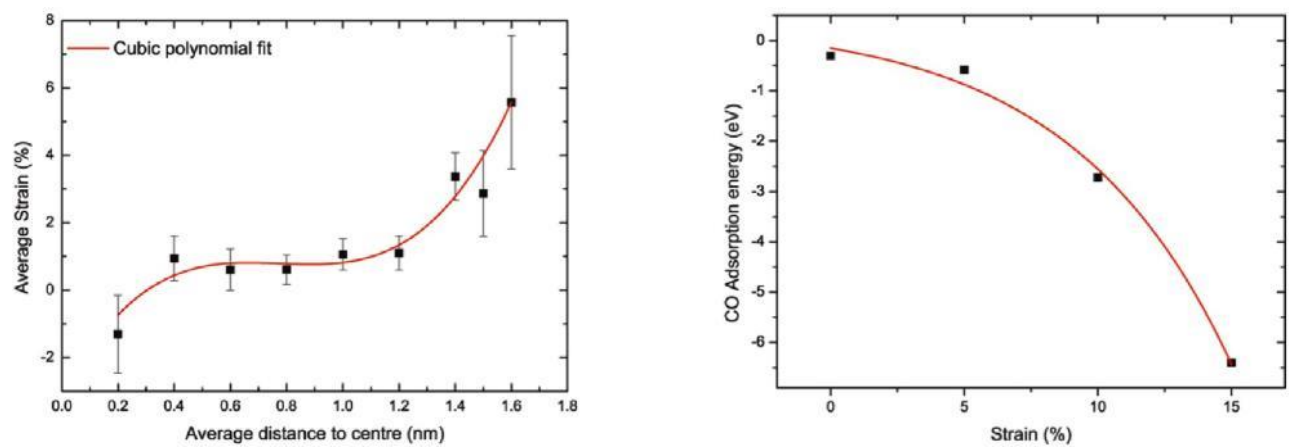

Figure 3. Distribution of strain measured as a function of distance from the center of a decahedral gold nanoparticle.

As a basis for comparison, a single crystal gold nanoparticle in the [110] orientation is shown in Figure 1c. This particle is again as deposited and prepared and imaged under the same conditions as the decahedral particle in Figure 1a. Here the single crystal structure should have no inherent strain, therefore any strain measured will be a consequence of beam-induced atomic rearrangements, an artifact of the analysis or the effect of electronic density redistribution due to surface stress and high surface to volume ratio of the nanoparticle.[29] Analysis of the single crystal structure in Figure 1d shows the particle is largely unstrained with a Gaussian distribution of nearest neighbor distances and a mean of 0.286 ± 0.001 nm. Again these strains are relative to the bulk nearest neighbor distance (0.288 nm), and some small contraction relative to this value can be expected at this size range.[30] The surfaces of the single crystal particle are generally unstrained or slightly compressed, while the few areas that show expansion at the surface set an upper limit on nonstructure specific strain and apparent strain induced by the imaging or analysis process. However, by comparing the strain maps in Figure 1 panels b and d it is clear that the single crystal structure exhibits nothing like the strain of the crystallographically forbidden decahedral particle, and that the surface of the decahedral particle is highly strained compared to the surface of the single crystal particle.

In order to quantify the effect of such large surface strains intrinsic to the decahedral structure, density functional theory has been used to calculate the adsorption energy for CO on Au(111) surfaces with strain increasing up to 15%. The c(4 × 8) (111) surface cell was expanded in each direction of the (111) plane and then optimized in the direction normal to the (111) plane. The CO adsorption energies on the strained-Au(111) slabs were evaluated under similar calculating conditions to the free-strain surfaces. Full details of the DFT methodology can be found in the Supporting Information.

For the unstrained Au(111) surface the CO adsorption energy was found to be −0.31 eV, which is in good agreement with previous reports.[9] Upon the introduction of 5% strain, the adsorption energy increases in magnitude to −0.58 eV, while increasing the strain toward the larger amounts observed at the decahedral surface causes the magnitude of the adsorption energy to increase exponentially, as shown in Figure 4. At 10% strain, the adsorption energy is −2.72 eV and reaches −6.40 eV by 15%. Upon adsorption on the heavily strained surfaces the calculations predict significant atomic rearrangement with nearest neighbor distances contracting toward the bulk values as shown in Supporting Information Figure S6. Subsequent adsorption energies calculated on these restructured surfaces are significantly reduced and are comparable to those calculated for 5% strain. However, the calculations are performed on a model slab, free from the structural constraints experienced by the decahedral particle. Furthermore, these values are based on a surface expanded in all directions, while the expansion of interatomic distances observed is inhomogeneous and surface atoms may be heavily strained in relation to one nearest neighbor but less so to another. A detailed theoretical study including the fixed twin boundaries and variable interatomic distances measured is beyond the scope of the current paper. However, after binding to the strained active site it can be expected that adsorption may act as a trigger for surface atom movement and possibly internal restructuring at higher temperature and pressures.

Figure 4. DFT calculations of the effect of strain on the adsorption energy for carbon monoxide.

It seems clear that inducing a strain to a metal surface can significantly enhance the catalytic performance of a material. Previous studies seeking to explain the surprising activity of small gold nanoparticles have largely overlooked the effect of strain, and the only source to be considered has been the relatively small effect of support-induced strain. The multiply twinned icosahedral and decahedral structures are regularly observed and are expected to be the most stable forms of gold nanoparticles in the size range for which gold is chemically active. However, the nature of their crystallographically forbidden, non-space filling structure has been overlooked until now when considering the origin of the catalytic activity of small gold nanoparticles. Here, we have demonstrated the substantial inherent strain in a decahedral Au particle in the size range for which gold is chemically active. The strain is found to be greatest at the particle's surface with an average surface strain of 5.6% and many instances of 10% or greater surface expansion. DFT calculations show that such surface strain will have a huge impact on the adsorption energy of CO and therefore activity for carbon monoxide oxidation. We believe these results provide further explanation of the dramatic increase in catalytic activity of gold nanoparticles in this size range and further highlight the structural sensitivity of their properties.

## ASSOCIATED CONTENT

### *s Supporting Information

Further details of the sample preparation and microscopy are provided. Results from the geometrical phase analysis are given, as well as further results from the real space strain analysis approach. Finally, details of the DFT calculations are presented and a video file showing restructuring upon adsorption is included. This material is available free of charge via the Internet at http://pubs.acs.org.

## AUTHOR INFORMATION

### Corresponding Author
*E-mail: pratibha.gai@york.ac.uk.

### Notes
The authors declare no competing financial interest.

## ACKNOWLEDGMENTS

Johnson Matthey Plc and EPSRC are thanked for studentship (M.J.W.) and Professor A. Howie for helpful discussions.